\documentclass[aps,prc,preprint,floatfix]{revtex4}

\usepackage{amsmath}
\usepackage{graphicx}
\usepackage{bm}
\usepackage{color}
\newcommand{\eps}{\varepsilon}

\preprint{JLAB-THY-12-1664}

\begin{document}

\title{Pion momentum distributions in the nucleon \\
	in chiral effective theory}

\author{M.~Burkardt$^1$, K.~S.~Hendricks$^2$, Chueng-Ryong~Ji$^2$,
	W.~Melnitchouk$^3$, A.~W.~Thomas$^4$}
\affiliation{
    $^1$Department of Physics, New Mexico State University,
	Las Cruces, New Mexico 88003, USA	\\
    $^2$Department of Physics, North Carolina State University,
	Raleigh, North Carolina 27692, USA	\\
    $^3$\mbox{Jefferson Lab, 12000 Jefferson Avenue,
	Newport News, Virginia 23606, USA}	\\
    $^4$ARC Centre of Excellence in Particle Physics at the Terascale
	and CSSM, School of Chemistry and Physics,
	University of Adelaide, Adelaide SA 5005, Australia}

\date{\today}

\begin{abstract}
We compute the light-cone momentum distributions of pions in the nucleon
in chiral effective theory using both pseudovector and pseudoscalar
pion--nucleon couplings.  For the pseudovector coupling we identify
$\delta$-function contributions associated with end-point singularities
arising from the pion--nucleon rainbow diagrams, as well as from pion
bubble and tadpole diagrams which are not present in the pseudoscalar
model.  Gauge invariance is demonstrated, to all orders in the pion
mass, with the inclusion of Kroll-Ruderman couplings involving operator
insertions at the $\pi NN$ vertex.  The results pave the way for
phenomenological applications of pion cloud models that are manifestly
consistent with the chiral symmetry properties of QCD.
\end{abstract}

\maketitle

%%%%%%%%%%%%%%%%%%%%%%%%%%%%%%%%%%%%%%%%%%%%%%%%%%%%%%%%%%%%%%%%%%%%%%%%%
\section{Introduction}
\label{sec:intro}

Contributions from the meson cloud of the nucleon to deep-inelastic
scattering (DIS) were first discussed in the early 1970s by Drell,
Levy and Yan \cite{DLY70} and Sullivan \cite{Sul72}.  It was of little
interest, however, until in the early 1980s a possible enhancement of
the pion cloud of a bound nucleon was suggested as an explanation of
the nuclear EMC effect \cite{Eri83, Lle83}.  While that issue is still
of interest, the modern importance of the pion cloud in high-energy
scattering originates in the work of Thomas \cite{Tho83}, who showed
(within a particular chiral quark model) that the pion cloud
contribution provided a natural explanation for the observed
excess of non-strange over strange sea quarks in the nucleon.
This calculation also predicted an excess of $\bar{d}$ over $\bar{u}$
quarks in the nucleon sea, which was confirmed by the violation of the
Gottfried sum rule observed in the NMC experiment at CERN \cite{NMC},
and the $\bar d/\bar u$ ratio measured in the Drell-Yan reaction in
$pd$ and $pp$ scattering by the E866/NuSea collaboration at Fermilab
\cite{E866}.

Although this excess of $\bar{d}$ over $\bar{u}$, as well as the
$s-\bar{s}$ asymmetry \cite{Sig87}, was motivated by the physics of
spontaneous chiral symmetry breaking in QCD, the early calculations
were based on models whose connections to QCD were not manifest.
It was later realized \cite{TMS00}, however, that because of their
origin in chiral loops, these contributions had a nonanalytic
dependence on quark mass that could only be generated by a Goldstone
mechanism, which placed these effects on a more rigorous theoretical
footing.  In fact, it is a model-independent consequence of
spontaneous chiral symmetry breaking in QCD that $\bar{d} - \bar{u}$
and $s - \bar{s}$ are nonzero.  The only open question is how large
these asymmetries actually are.

This issue has become even more important in the last few years because
of the widespread interest in the so-called five-quark components of
baryon wave functions \cite{Vog00}.  In particular, there is some
evidence that the Roper resonance may not be a simple excitation of
a single valence quark in the nucleon but rather involves a large
five-quark contribution \cite{Kre00, Mel02, Gua04, Mat05, Mah09}.
This naturally leads to questions about such components in the wave
function of the nucleon, and there have been suggestions that these
may be sizable.  Even though the pentaquark now appears to be defunct,
legitimate questions about five-quark components in baryon spectroscopy
remain, especially within models such as the chiral quark soliton model
\cite{Goe11}.

In the context of low energy tests of the Standard Model this issue
is also very topical, with a potential asymmetry between the $s$ and 
$\bar{s}$ parton distributions potentially yielding a large correction
to the value of $\sin^2\theta_W$ derived from the NuTeV measurement
\cite{Mas07, Ben10}.  Model-dependent estimates of the five-quark
component of the nucleon wave function involving three light quarks plus
$s \bar{s}$ pairs were recently used to investigate this correction
\cite{Alw04, Wak05, Din05, Cha11}.

In all of these examples, the contribution from the meson cloud of the
nucleon has been a crucial factor in reconciling the physics results.
Indeed, because of the model-independent constraints imposed through
chiral symmetry on the nonanalytic behavior \cite{TMS00, LiP71, Gas87},
the presence of meson cloud contributions to the various observables
is firmly established, justifying the extensive theoretical attention
that has been paid to this issue over the past decade.
(We should note, however, that in chiral perturbation theory
observables in general receive contributions from pions as well
as local short-distance operators, or counter-terms, and it is
possible under renormalization to move strength between them
\cite{Mei07}; the nonanalytic contributions, on the other hand,
remain model independent.)
A further motivation relates to the fact that lattice QCD is now
providing considerable information on the moments of parton
distribution functions \cite{Bal12}, but at larger quark masses
than occur in nature.  The extrapolation of those moments as
a function of quark mass to the physical point is critically
dependent on knowing the correct nonanalytic behavior \cite{Det01},
and that in turn is uniquely determined by the pion cloud of the
nucleon.

Given the phenomenological importance of the meson cloud, it is
unfortunate that the literature contains a number of sometimes
contradictory results for meson cloud contributions to DIS.
In particular, the majority of calculations, from the original
work of Drell {\em et al.} \cite{DLY70} and Sullivan \cite{Sul72},
to the early studies of the nonanalytic behavior \cite{TMS00},
and essentially all model analyses of pion cloud effects in DIS
in between \cite{Spe97}, have used pseudoscalar (PS) coupling,
which is, by itself, inconsistent with chiral symmetry
\cite{Gel60, Gel62, Wei67}.  The restoration of chiral symmetry
can be achieved through the addition of a scalar ``$\sigma$'' field;
however, determining its practical consequences for DIS is problematic
because of uncertainties in identifying the nature of low-mass scalars
in meson spectroscopy \cite{Pen07}.

More recently, the matrix elements of nonsinglet twist-2 operators
were computed by Chen and Ji \cite{Che01, Che02} and Arndt and Savage
\cite{Arn02} using lowest order, heavy baryon chiral perturbation
theory.  By construction this theory uses pseudovector (PV) $\pi N$
couplings, which are manifestly invariant under chiral transformations
\cite{Wei67, Jen91, BKM95}.  The twist-2 matrix elements are related
through the operator product expansion to moments of parton
distributions measured in inclusive DIS.
At lowest order in the low-energy expansion (minimum number of
derivatives of the pion field $\bm{\pi}$), the effective chiral
Lagrangian for the interaction of pions and nucleons, consistent with
chiral symmetry, can be written as \cite{Wei67, Jen91, BKM95, Kra90}
\begin{eqnarray}
{\cal L}_{\pi N}
&=& {g_A \over 2 f_\pi}\,
    \bar\psi_N \gamma^\mu \gamma_5\,
    \bm{\tau} \cdot \partial_\mu \bm{\pi}\, \psi_N\
 -\ {1 \over (2 f_\pi)^2}\,
    \bar\psi_N \gamma^\mu\, \bm{\tau} \cdot
    (\bm{\pi} \times \partial_\mu \bm{\pi})\, \psi_N,
\label{eq:LpiN}
\end{eqnarray}
where $\psi_N$ is the nucleon field, $f_\pi = 93$~MeV is the pion
decay constant, and $g_A = 1.267$ is the nucleon axial vector charge.
The first term in the Lagrangian (\ref{eq:LpiN}) gives rise to the
well-known ``rainbow'' diagram in which a pion is emitted and
reabsorbed by the nucleon at different space-time points.
The second is the so-called Weinberg-Tomozawa term \cite{Wei67, Tom66},
in which two pion fields couple to the nucleon at the same point,
and gives the leading contribution to S-wave pion--nucleon scattering
\cite{Wei66}.
It also generates the pion tadpole or bubble diagrams, which in the
presence of external fields generally give non-vanishing contributions
to nucleon matrix elements.
This term does not appear in the PS theory, which leads to some of
the differences between the moments of twist-2 parton distributions
from pion loops computed in the PS \cite{TMS00} and PV
\cite{Che01, Che02, Arn02} theories.

In this paper we present a detailed analysis of the light-cone
momentum distributions of pions, and the corresponding recoil
baryons, inside a physical nucleon within the chirally symmetric
effective field theory defined by the Lagrangian (\ref{eq:LpiN}).
We further contrast the results with those obtained in the PS
theory, conventionally defined by the Lagrangian
\begin{eqnarray}
{\cal L}^{\rm PS}_{\pi N}
&=& -g_{\pi NN}\,
    \bar\psi_N\, i \gamma_5\, \bm{\tau} \cdot \bm{\pi}\, \psi_N,
\label{eq:Lps}
\end{eqnarray}
where the $\pi NN$ coupling constant $g_{\pi NN}$ is related to
$g_A$ and $f_\pi$ by the Goldberger-Treiman relation \cite{GT58},
\begin{eqnarray}
{g_{\pi NN} \over M} &=& {g_A \over f_\pi}\, ,
\label{eq:GT}
\end{eqnarray}
with $M$ the nucleon mass.
Unlike the earlier chiral effective theory calculations which only
computed the light-cone distributions of pions \cite{Che02} or
considered the nonanalytic behavior of their moments to lowest
order in the pion mass $m_\pi$ \cite{Che01, Arn02}, we compute
the complete set of diagrams relevant for DIS from nucleons
dressed by pions resulting from the Lagrangians (\ref{eq:LpiN})
and (\ref{eq:Lps}), without taking the heavy baryon limit.
In particular, we demonstrate explicitly the consistency of the
computed distribution functions with electromagnetic gauge invariance.
This requires consideration of the Kroll-Ruderman terms \cite{Kro54},
which although entering at higher orders in $m_\pi$, are nonetheless
essential for ensuring conservation of charge to all orders in $m_\pi$
\cite{Dre92}.

Formally, the light-cone distribution functions can be defined by
considering the vertex renormalization constant $Z_1^{-1} - 1$
for the physical $\gamma^* NN$ vertex.  If the contributions from
the pion cloud are not large, then $Z_1 \approx 1$ and one has
		$Z_1^{-1} - 1 \approx 1 - Z_1$.
The light-cone distributions $f_i(y)$ associated with a particular
contribution $i$ can then be defined as \cite{JMT_Z1} 
\begin{eqnarray}
(1 - Z_1)_i &=& \int_0^1\!dy\, f_i(y),
\label{eq:fydef}
\end{eqnarray}
where $y = k_+/p_+$ is the momentum fraction of the physical nucleon
carried by the pion, with $k$ and $p$ the four-momenta of the pion
and physical nucleon, respectively.
Note that in this work we define the light-front ``$+$'' and ``$-$''
components of a four-vector $v^\mu$ as $v^\pm = v^0 \pm v^z$. 
It will be most natural to perform the calculation in light-front
coordinates; however, as demonstrated in Ref.~\cite{JMT_Sig} for
the nucleon self-energy, the results for the model-independent,
long-distance physics associated with the pion cloud are reproduced
in any formalism, be it instant form, covariant, or light-front.

Because the tadpole or bubble diagrams in the PV theory involve
pions emitted and absorbed at the same point, these can only contribute
at zero light-cone momentum fraction, $y=0$.  A novel feature of the PV
theory, however, is the appearance of $\delta$-function contributions
arising also from the rainbow diagram at $y=0$.  Inclusion of these
singular terms is in fact vital for preserving gauge invariance in
the PV theory.  In the following we will elucidate the origin of the
$\delta$-function terms and discuss their possible physical implications.

The rest of this paper is organized as follows.
We derive in Sec.~\ref{sec:fypi} the light-cone momentum distribution
functions for pions in a physical nucleon, including both the pion
rainbow and bubble contributions.
The analogous nucleon momentum distributions are discussed in
Sec.~\ref{sec:fyN}, where the Kroll-Ruderman couplings required
by gauge invariance are introduced.  The complete set of pion and
nucleon distributions allows us for the first time to explicitly
verify the symmetry relations between them.
The application of the results to inclusive DIS is described in
Sec.~\ref{sec:sumrules} within the convolution approach, where we
discuss the effect of the pion loop corrections, including the
$\delta$-function contributions, on various sum rules.
Finally, we summarize our findings and their implications
in Sec.~\ref{sec:conc}.

%%%%%%%%%%%%%%%%%%%%%%%%%%%%%%%%%%%%%%%%%%%%%%%%%%%%%%%%%%%%%%%%%%%%%%%%%
\section{Pion light-cone momentum distributions}
\label{sec:fypi}

\begin{figure}[ht]
\includegraphics[width=9cm]{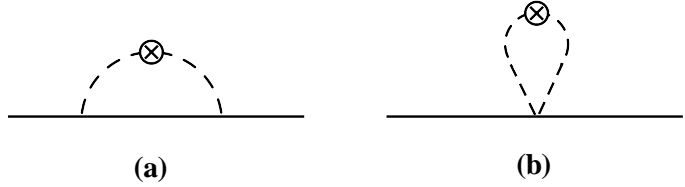}
\caption{Contributions to the pion light-cone momentum distributions
	in the nucleon, from (a) the pion rainbow, and (b) the pion
	bubble diagrams.}
\label{fig:pi}
\end{figure}

Assuming isospin symmetry, the contributions from pion loops
to nucleon matrix elements cancel for isoscalar combinations.
In this work we therefore consider only isovector light-cone
momentum distributions, $f_i^{(iv)}(y) \equiv f_i^p(y) - f_i^n(y)$.
The distributions are evaluated by taking the ``$+$'' components
of the operators.  We begin by considering the contributions from
diagrams involving a direct coupling to the pion, as illustrated
in Fig.~\ref{fig:pi}.

The isovector light-cone momentum distribution function for the
pion rainbow diagram in Fig.~\ref{fig:pi}(a) is given by
\begin{eqnarray}
\hspace*{-0.3cm}
f_\pi^{(iv)}(y)
&=& 4M \left( {g_A \over 2 f_\pi} \right)^2
    \int\!\!{d^4k \over (2\pi)^4}
    \bar u(p)\, 
    (k\!\!\!\slash \gamma_5)
    {i (p\!\!\!\slash - k\!\!\!\slash + M) \over D_N}\,
    (\gamma_5 k\!\!\!\slash)
    u(p)
    {i \over D_\pi} {i \over D_\pi}\, 2k^+ \delta(k^+ - y p^+),\nonumber\\
& &
\label{eq:fypi_def}
\end{eqnarray}
where the pion and nucleon propagators are given by
\begin{eqnarray}
D_\pi &\equiv& k^2 - m_\pi^2 + i\eps,\ \ \ \ \ \
D_N\   \equiv\ (p-k)^2 - M^2 + i\eps,
\end{eqnarray}
respectively.  In Eq.~(\ref{eq:fypi_def}) and throughout this work
the spinors $u(p)$ are normalized such that $\bar u(p)\, u(p) = 1$.
(Note that for the pion coupling diagrams, the isovector combination
$p-n$ is equivalent to the notation ``$\pi^+ - \pi^-$'' \cite{Che02}.)
Using the Dirac equation, the distribution function in
Eq.~(\ref{eq:fypi_def}) can be decomposed into several terms,
\begin{eqnarray}
f_\pi^{(iv)}(y)
&=& -4i \left( {g_A \over 2 f_\pi} \right)^2
    \int\!\!{d^4k \over (2\pi)^4}
    \left[ {4M^2\, p\cdot k \over D_\pi^2 D_N}\
	 +\ {2M^2 \over D_\pi^2}
	 +\ {p \cdot k \over D_\pi^2}
    \right]
    2y\, \delta\left(y - {k^+\over p^+}\right).
\label{eq:fypi_D}
\end{eqnarray}
To evaluate the expression in Eq.~(\ref{eq:fypi_D}), we proceed
to first integrate over the $k^-$ component of the pion momentum.
Without loss of generality, we work in a frame where $p_\perp=0$.
For the first term in (\ref{eq:fypi_D}) we use Cauchy's integral
formula and take the single nucleon pole by closing the contour
in the upper half-plane,
\begin{eqnarray}
D_N &=& (p^+ - k^+)
    \left( p^- - k^- - {k_\perp^2 + M^2 + i\eps \over p^+ - k^+}
    \right)\
\longrightarrow\ 0.
\end{eqnarray}
Note that the arc contribution at infinity vanishes for this term.
A straightforward calculation then yields the contribution which we
associate with the on-shell part of the nucleon propagator,
\begin{eqnarray}
f^{\rm(on)}(y)  
&=& {g_A^2 M^2 \over (4\pi f_\pi)^2}
    \int\!dk_\perp^2\,
    { y (k_\perp^2 + y^2 M^2) \over
     \left[ k_\perp^2 + y^2 M^2 + (1-y) m_\pi^2 \right]^2 },
\label{eq:fyon}
\end{eqnarray} 
where for convenience we have factored out all isospin factors.
This term corresponds to the usual pion distribution function
for the ``Sullivan process'' \cite{Sul72}, which is computed
from a PS $\pi N$ coupling.
Performing the $k_\perp$ integration with an ultraviolet cut-off
$\Lambda$ gives
\begin{eqnarray}
f^{\rm(on)}(y)
%
% &=& -{g_A^2 M^2 \over (4\pi f_\pi)^2}\, y
% \left[
%    \frac{\Lambda^2 m_\pi^2 (1-y)}
%	{(\Lambda^2 + m_\pi^2 (1-y) + M^2 y^2)(m_\pi^2 (1-y) + M^2 y^2)}\
% \right.						\nonumber\\
% & & \hspace*{2cm}
% \left.
% +\ \log\left( \frac{m_\pi^2 (1-y) + M^2 y^2}
%		   {\Lambda^2 + m_\pi^2 (1-y) + M^2 y^2}
%       \right)
% \right]						\nonumber\\
%
&=& -{g_A^2 M^2 \over (4\pi f_\pi)^2}\, y
\left[
   \frac{m_\pi^2 (1-y)}{m_\pi^2 (1-y) + M^2 y^2}\
+\ \log\left( \frac{m_\pi^2 (1-y) + M^2 y^2}{\Lambda^2} \right)
\right]
+ {\cal O}\left( \frac{M^2}{\Lambda^2} \right),		\nonumber\\
& &
\label{eq:fyon_final}
\end{eqnarray}
where only the leading term in $\Lambda$ has been kept.

In the second and third terms in Eq.~(\ref{eq:fypi_D}),
the cancellation of the nucleon propagators $D_N$ leaves
tadpole-like contributions involving pion propagators only.
This will have important consequences for the structure and
interpretation of the distribution functions.
Integrating the second term in (\ref{eq:fypi_D}) proportional to
$y/D_\pi^2$ over $k^-$ gives a result which is zero everywhere
except at $k^+ = y p^+ = 0$,
\begin{eqnarray}
\int dk^- {1 \over D_\pi^2}
&=& {2\pi i \over k_\perp^2 + m_\pi^2}\, \delta(k^+).
\label{eq:intDpi}
\end{eqnarray}
After multiplication by $y$, this term therefore vanishes.

Evaluation of the third term in Eq.~(\ref{eq:fypi_D}) proportional
to $(y\, p \cdot k) /D_\pi^2$ requires particular care.
Since $p \cdot k = (p^+ k^- + p^- k^+)/2$, integration over $k^-$
involves terms such as in (\ref{eq:intDpi}) which vanish, as well as
terms of the type $\int dk^-\, (k^-/D_\pi^2)$.
To evaluate the latter, one can make use of the identity
\begin{eqnarray}
{\ k^- \over D_\pi^2}
&=& -{\partial \over \partial k^+} {1 \over D_\pi}
\end{eqnarray}
to write
\begin{eqnarray}
\int dk^- {k^+ k^- \over D_\pi^2}
% &=& -k^+ {\partial \over \partial k^+} \int dk^- {1 \over D_\pi}\
&=& \int dk^- {1 \over D_\pi}\
 =\ 2\pi i\, \log\left({k_\perp^2 + m_\pi^2 \over \mu^2}\right)\,
    \delta(k^+),
\label{eq:Dpi_id}
\end{eqnarray}
where $\mu$ is a mass parameter.
This gives rise to a $\delta$-function contribution (again defined
without isospin factors)
\begin{eqnarray}
f^{(\delta)}(y)
&=& {g_A^2 \over 4 (4\pi f_\pi)^2}
    \int\!dk_\perp^2\,
    \log\left({k_\perp^2 + m_\pi^2 \over \mu^2}\right)\,
    \delta(y),
\label{eq:fydelta}
\end{eqnarray}
with the total pion distribution function then given by
\begin{eqnarray}
f_\pi^{(iv)}(y) &=& 4 f^{\rm(on)}(y) + 4 f^{(\delta)}(y).
\label{eq:fypi}
\end{eqnarray}
When integrated over $k_\perp$ from 0 to $\Lambda$, the function
$f^{(\delta)}(y)$ becomes
\begin{eqnarray}
f^{(\delta)}(y)
%
% &=& {g_A^2 \over 4 (4\pi f_\pi)^2}
%    \left[ - \Lambda^2
%           + \Lambda^2 \log\left( \frac{\Lambda^2+m_\pi^2}{\mu^2} \right)
%           - m_\pi^2 \log\left( \frac{m_\pi^2}{\Lambda^2+m_\pi^2} \right)
%    \right]                                     \nonumber\\
%
&=& -{g_A^2 \over 4 (4\pi f_\pi)^2}
    m_\pi^2 \log m_\pi^2\ \delta(y)\
 +\ \textrm{terms involving $\Lambda$},
\label{eq:fydelta_final}
\end{eqnarray}
where we have isolated the leading nonanalytic term in $m_\pi^2$,
whose coefficient is independent of the short-distance physics
parametrized by $\Lambda$.

Thus the pion rainbow diagram has contributions from the regular
$y > 0$ region as well as the ultrasoft $y = 0$ point.
In the PS formulation of pion--nucleon interactions, where the pion
light-cone momentum distribution is given by just the first term in
Eq.~(\ref{eq:fypi_D}), the $f^{(\delta)}$ term in (\ref{eq:fypi}) does
not arise, and the total PS pion distribution is given by \cite{DLY70}
\begin{eqnarray}
f_{\pi ({\rm PS})}^{(iv)}(y) &=& 4 f^{\rm(on)}(y).
\end{eqnarray}

In addition to the pion rainbow diagram in Fig.~\ref{fig:pi}(a),
there is also a contribution from the term in the Lagrangian
(\ref{eq:LpiN}) involving two pion fields, which gives rise
to the bubble diagram in Fig.~\ref{fig:pi}(b).
A straightforward calculation of the resulting contribution
to the pion light-cone momentum distribution gives
\begin{eqnarray}
f_{\pi \rm (bub)}^{(iv)}(y)
&=& {M \over f_\pi^2}
    \int\!\!{d^4k \over (2\pi)^4}\,
    \bar u(p)\, (-i k\!\!\!\slash)\, u(p)\, 
    {i \over D_\pi} {i \over D_\pi}\, 2k^+ \delta(k^+ - y p^+)	\\
&=& {i \over f_\pi^2}
    \int\!\!{d^4k \over (2\pi)^4}\,
    {p \cdot k \over D_\pi^2}\,
    2y\, \delta\left(y - {k^+\over p^+}\right).
\label{eq:fypitad_def}
\end{eqnarray}
A calculation similar to that for the $f^{(\delta)}(y)$ term above
gives the final pion bubble contribution as
\begin{eqnarray}
f_{\pi \rm (bub)}^{(iv)}(y)
&=& -{1 \over (4\pi f_\pi)^2}
    \int\!dk_\perp^2\,
    \log\left( {k_\perp^2+m_\pi^2 \over \mu^2} \right) \delta(y)\
\equiv\ 2 f^{\rm (bub)}(y),
\label{eq:fypitad}
\end{eqnarray}
where for convenience we have defined the distribution
$f^{\rm (bub)}(y)$ without the isospin factor 2.
Integrating over $k_\perp$ from 0 to $\Lambda$ then yields a result
for $f^{\rm (bub)}(y)$ similar to that for the $f^{(\delta)}(y)$
term in Eq.~(\ref{eq:fydelta}),
\begin{eqnarray}
f^{\rm (bub)}(y)
&=& {1 \over 2 (4\pi f_\pi)^2}
    m_\pi^2 \log m_\pi^2\ \delta(y)\
 +\ \textrm{terms involving $\Lambda$}		\\
&=& -{2 \over g_A^2}\, f^{(\delta)}(y).
\label{eq:fytad_final}
\end{eqnarray}
Finally, combining the results in this section, the total isovector
pion distribution in the nucleon, including both the pion rainbow and
bubble diagrams in Fig.~\ref{fig:pi}, can be written as
\begin{eqnarray}
f_\pi^{(iv)}(y)\ +\ f_{\pi \rm (bub)}^{(iv)}(y)
&=& -{4 g_A^2 M^2 \over (4\pi f_\pi)^2}\, y
    \left[
      \frac{m_\pi^2 (1-y)}{m_\pi^2 (1-y) + M^2 y^2}\
      +\ \log\left( m_\pi^2 (1-y) + M^2 y^2 \right)
    \right]					\nonumber\\
& &
 +\ {(1-g_A^2) \over (4\pi f_\pi)^2}
    m_\pi^2 \log m_\pi^2\ \delta(y)\
 +\ \textrm{terms involving $\Lambda$},
\label{eq:fypi_final}
\end{eqnarray}
where we have explicitly separated the long-distance contributions
from the short-distance effects involving the cut-off $\Lambda$.
In particular, the total term proportional to the $\delta$-function
is found to vanish in the limit $g_A \to 1$, as one would have in
a PS pion--nucleon theory \cite{Che01, Che02}.
Note that a similar behavior for the $\delta(y)$ term is observed
for the pion cloud contribution to the isovector form factor
\cite{Str10, Kai03}.

%%%%%%%%%%%%%%%%%%%%%%%%%%%%%%%%%%%%%%%%%%%%%%%%%%%%%%%%%%%%%%%%%%%%%%%%%
\section{Nucleon light-cone distributions}     
\label{sec:fyN}

\begin{figure}[t]
\includegraphics[width=9cm]{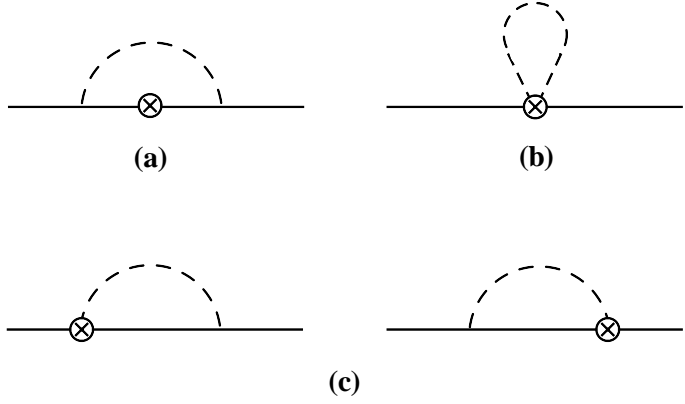}
\caption{Contributions to the nucleon light-cone momentum
	distributions in the nucleon, from
	(a) the nucleon rainbow, and
	(b) the tadpole diagrams, as well as
	(c) the contributions from the Kroll-Ruderman
	coupling required by gauge invariance.}
\label{fig:N}
\end{figure}

For the nucleon rainbow diagram in Fig.~\ref{fig:N}(a),
the distribution function for the neutron is $2\, \times$ that
of the proton, so that for the isovector distribution one has
$f_N^{(iv)}(y) = -f_p(y)$.  Specifically, the isovector nucleon
distribution function is given by
\begin{eqnarray}
f_N^{(iv)}(y)
&=& -M \left({g_A \over 2f_\pi}\right)^2
    \int\!\!{d^4k \over (2\pi)^4}\,
    \bar u(p)\,
    (k\!\!\!\slash \gamma_5)\,
    {i (p\!\!\!\slash - k\!\!\!\slash + M) \over D_N}\,
    \gamma^+\,
    {i (p\!\!\!\slash - k\!\!\!\slash + M) \over D_N}\,
    (\gamma_5 k\!\!\!\slash)\,
    u(p)					\nonumber\\
& & \hspace{4cm} \times
    {i \over D_\pi}\,
    \delta(k^+ - y p^+),
\end{eqnarray}
which, using the Dirac equation, can be written as
\begin{eqnarray}
\hspace*{-0.5cm}
f_N^{(iv)}(y)
&=& i \left( {g_A \over 2 f_\pi} \right)^2
    \int\!\!{d^4k \over (2\pi)^4}
    \left[ {4 M^2 (k^2 - 2y\, p\cdot k) \over D_\pi D_N^2}\
	 -\ {4 M^2 y \over D_\pi D_N}\
	 -\ {1 \over D_\pi}
    \right]
    \delta\left(y - {k^+\over p^+}\right).
\label{eq:fyN_def}
\end{eqnarray}
The first term ($\sim 1/D_\pi D_N^2)$ in Eq.~(\ref{eq:fyN_def})
corresponds to the on-shell nucleon contribution, and is proportional
to the function $f^{\rm(on)}(y)$ in Eq.~(\ref{eq:fyon}).
The second term ($\sim 1/D_\pi D_N)$ arises from the off-shell
components of the nucleon propagator, while the third term involving
the single pion propagator ($\sim 1/D_\pi$) contributes only at $y=0$,
and is proportional to $f^{(\delta)}(y)$.
The total isovector nucleon light-cone distribution function arising
from the nucleon rainbow diagram is then given by
\begin{eqnarray}
f_N^{(iv)}(y) &=& -f^{\rm(on)}(y) - f^{\rm(off)}(y) + f^{(\delta)}(y),
\label{eq:fyN}
\end{eqnarray}
where
\begin{eqnarray}
f^{\rm(off)}(y)
&=& - {g_A^2 M^2 \over (4\pi f_\pi)^2}
    \int\!dk_\perp^2\,
    { y \over k_\perp^2 + y^2 M^2 + (1-y) m_\pi^2 }	\\
%
% &=& {g_A^2 M^2 \over (4\pi f_\pi)^2}\, y
%    \log\left[ \frac{m_\pi^2 (1-y) + M^2 y^2}
%		    {\Lambda^2 + m_\pi^2 (1-y) + M^2 y^2}
%	\right]						\nonumber\\
%
& &							\nonumber\\
&=& {g_A^2 M^2 \over (4\pi f_\pi)^2}\, y
    \log\left( \frac{m_\pi^2 (1-y) + M^2 y^2}
                    {\Lambda^2}
        \right)
+ {\cal O}\left( \frac{M^2}{\Lambda^2} \right),
\label{eq:foff}
\end{eqnarray}
with the $k_\perp$ integration taken up to the cut-off scale $\Lambda$.

For the pseudoscalar model the nucleon light-cone momentum distribution
is again just given by the on-shell term in Eq.~(\ref{eq:fyN}),
\begin{eqnarray}
f_{N ({\rm PS})}^{(iv)}(y) &=& -f^{\rm(on)}(y).
\end{eqnarray}
This result was also obtained in the infinite momentum frame
calculation of Drell, Levy and Yan \cite{DLY70}.

On the other hand, the PV theory contains, in addition to the off-shell
and $\delta$-function pieces, the contribution from the operator
insertion at the $NN\pi\pi$ vertex, Fig.~\ref{fig:N}(b).
The distribution function associated with this diagram can be written
\begin{eqnarray}
f_{N \rm (tad)}^{(iv)}(y)
&=& -{M \over f_\pi^2}
    \int\!\!{d^4k \over (2\pi)^4}\,
    \bar u(p)\, \gamma^+\, u(p)\, 
    {i \over D_\pi}\, \delta(k^+ - y p^+)	\\
&=& -{i \over f_\pi^2}
    \int\!\!{d^4k \over (2\pi)^4}\,
    {1 \over D_\pi}\,
    \delta\left(y - {k^+\over p^+}\right).
\label{eq:fyNtad_def}
\end{eqnarray}
Performing the $k^-$ integration and again using the relation
in Eq.~(\ref{eq:Dpi_id}) then gives
\begin{eqnarray}
f_{N \rm (tad)}^{(iv)}(y)
&=& - 2 f^{\rm (bub)}(y).
\label{eq:fyNtad}
\end{eqnarray}
Comparing the pion bubble and tadpole contributions,
Eqs.~(\ref{eq:fypitad}) and (\ref{eq:fyNtad}), one finds that
their sum vanishes,
\begin{eqnarray}
f_{\pi \rm (bub)}^{(iv)}(y)\ +\ f_{N \rm (tad)}^{(iv)}(y) &=& 0.
\label{eq:tadsumrule}
\end{eqnarray}
These contributions themselves therefore have no net effect
on the sum of the pion and nucleon light-cone distributions,
and hence on the total nucleon charge, as will be discussed
in Sec.~\ref{sec:sumrules} below.

Finally, the light-cone momentum distribution associated with the
Kroll-Ruderman diagrams in Fig.~\ref{fig:N}(c), which arises from
the derivative coupling in the PV theory, is given by
\begin{eqnarray}
f_{\rm KR}^{(iv)}(y)
&=& 4M \left({g_A \over 2f_\pi}\right)^2
    \int\!\!{d^4k \over (2\pi)^4}\,
    \bar u(p)\, 
    \left[ k\!\!\!\slash \gamma_5
	   {i (p\!\!\!\slash - k\!\!\!\slash + M) \over D_N}
	   i \gamma^+ \gamma_5\
    \right.					\nonumber\\
& & \hspace{4cm}
    \left.
	+\ i \gamma_5 \gamma^+
	    {i (p\!\!\!\slash - k\!\!\!\slash + M) \over D_N}
	    \gamma_5 k\!\!\!\slash
    \right] 
    u(p)\,
    {i \over D_\pi}\,
    \delta\left(y - {k^+\over p^+}\right)	\\
&=& 4i \left( {g_A \over 2 f_\pi} \right)^2
    \int\!\!{d^4k \over (2\pi)^4}
    \left[ {4 M^2 y \over D_\pi D_N}\ +\ {2 \over D_\pi} \right]\,
    \delta\left(y - {k^+\over p^+}\right).
\label{eq:fyKR_def}
\end{eqnarray}   
After $k^-$ integration, one then obtains
\begin{eqnarray}
f_{\rm KR}^{(iv)}(y) &=& 4 f^{\rm (off)}(y)\ -\ 8 f^{(\delta)}(y).
\label{eq:fyKR}
\end{eqnarray}
One observes then that the sum of the pion [Eq.~(\ref{eq:fypi})]
and KR [Eq.~(\ref{eq:fyKR})] distributions is proportional to the
nucleon light-cone distribution (Eq.~(\ref{eq:fyN})),
\begin{eqnarray}
f_\pi^{(iv)}(y)\ +\ f_{\rm KR}^{(iv)}(y)
&=& 4 f^{\rm (on)}(y)\ +\ 4 f^{\rm (off)}(y)\ -\ 4 f^{(\delta)}(y)\
 =\ -4 f_N^{(iv)}(y),
\label{eq:cons}
\end{eqnarray}
which, as will be discussed in the next section, is a necessary
condition for the gauge invariance of the theory.

%%%%%%%%%%%%%%%%%%%%%%%%%%%%%%%%%%%%%%%%%%%%%%%%%%%%%%%%%%%%%%%%%%%%%%%%%
\section{Sum rules}
\label{sec:sumrules}

The effect of pion loops on observables such as parton distribution
functions can be computed by considering DIS from a nucleon viewed in
the infinite momentum frame \cite{DLY70}.  Here the total contribution
to the nucleon isovector quark distribution $q_N^{(iv)}$ from the pionic
corrections in Figs.~\ref{fig:pi} and \ref{fig:N} can be written in
convolution form (for a review see {\it e.g.} Ref.~\cite{Spe97}),
\begin{eqnarray}
q_N^{(iv)}(x)
&=& Z_2\, q_{N_0}^{(iv)}(x)\
 +\ \sum_i \Big( f_i^{(iv)} \otimes q_i^{(iv)} \Big)(x),
\label{eq:qNiv}
\end{eqnarray}
where $q_{N_0}^{(iv)}$ is the isovector quark distribution in the
bare nucleon state $N_0$, and the sum over $i$ includes the nucleon
and pion rainbow, bubble, tadpole and KR contributions.
The symbol $\otimes$ denotes convolution; for the pion rainbow
contribution, for instance, one has
$(f_\pi^{(iv)} \otimes q_\pi^{(iv)})(x)
= \int_x^1 (dy/y)\, f_\pi^{(iv)}(y)\, q_\pi^{(iv)}(x/y)$,
where $q_\pi^{(iv)}$ is the isovector quark distribution in the pion,
and similarly for the other terms.
While the shapes of the bare quark distribution functions,
and especially for the KR and tadpole terms, are not known
{\it a~priori}, charge conservation is nevertheless ensured
as long as the distributions are normalized according to
\begin{eqnarray}
\int_0^1 dx\, q_i^{(iv)}(x) &=& 1,
\end{eqnarray}
for all $i = N, \pi, {\rm KR}, \pi N$.

The wave function renormalization factor $Z_2$ in Eq.~(\ref{eq:qNiv})
can be evaluated from the nucleon self-energy $\Sigma$ by \cite{BjDr66,
Hec02, AM12, JMT_Comment},
\begin{eqnarray}
Z_2^{-1} - 1
&=& \left. { \partial \Sigma \over \partial p_0}
    \right|_{p_0=M},
\end{eqnarray}
where the derivative is taken at the nucleon pole, $p_0=M$.
For weak pion fields, explicit evaluation gives
\begin{eqnarray}
Z_2^{-1} - 1\ \approx\ 1 - Z_2
&=& 3 \int_0^1 dy\,
    \left( f^{\rm(on)}(y) + f^{\rm(off)}(y) - f^{(\delta)}(y) \right),
\label{eq:Z2}
\end{eqnarray}
where the factor 3 arises from the sum of the proton and neutron
intermediate state contributions, for either a proton or neutron
initial state.
In fact, Eq.~(\ref{eq:Z2}) is required by the Ward-Takahashi identity
which relates the wave function and vertex renormalization factors by
$Z_2 = 3 (1 - Z_1^N)$, where $Z_1^N$ is the lowest moment of the
nucleon distribution function $f_N$ for a {\it proton} initial state.

Integrating the convolution expressions in Eq.~(\ref{eq:qNiv}) over
all $x$ and using the relation in Eq.~(\ref{eq:cons}) one obtains,
for a nonsinglet quark distribution $q_N^{(iv)}(x)$ normalized to
unity, the quark number sum rule,
\begin{eqnarray}   
\int_0^1 dx\, q_N^{(iv)}(x)\ =\ 1
&=& Z_2\ +\ 3 \int_0^1 dy\,
      \left( f^{\rm(on)}(y) + f^{\rm(off)}(y) - f^{(\delta)}(y) \right),
\label{eq:norm}
\end{eqnarray}
with the contributions from the $f_{N \rm (tad)}^{(iv)}$
and $f_{\pi \rm (bub)}^{(iv)}$ terms canceling as in
Eq.~(\ref{eq:tadsumrule}).
{}From the expression for $Z_2$ in Eq.~(\ref{eq:Z2}) one can verify
that the baryon number of the nucleon is not modified by pion loops.

Inclusion of the $y=0$ contributions is vital for formal sum rules
to be satisfied; if these were to be evaluated for $y>0$ only, the
$\delta(y)$ terms would be missed.  Indeed, as was illustrated in
Ref.~\cite{Bur02} in the case of the Bukhardt--Cottingham sum rule
for the twist-3 polarized parton distribution $g_2(x)$, if such
$\delta(x)$ terms are present then sum rules based on the formal
operator product expansion appear to be violated when evaluated for
$x>0$ only.  The same was observed for the corresponding sum rule
for the chirally odd twist-3 parton distribution $h_2(x)$.
In the context of dispersion relations, $\delta(x)$ terms reflect the
presence of subtractions.  A deep-inelastic scattering experiment to
test such sum rules would thus find them to be violated, since $x=0$ is
never measured.  In the convolution calculation the $\delta(y)$ terms
would thus imply a $\delta(x)$ contribution to the quark distribution
$q_N^{(iv)}(x)$, which would lead to the violation of both the
Gottfried \cite{Got67} and Adler sum rules \cite{Adl65}, if these
are applied to nonzero $x$ only.

The appearance of the $\delta(x)$ terms in the isovector quark
distributions is possibly an artifact of taking chiral perturbation
theory too literally in this context.  For example, although a
bubble diagram in the chiral effective theory inevitably leads to
an $s$-independent contribution to the forward Compton amplitude ---
and hence a subtraction in the corresponding dispersion relation ---
it may no longer be justified to assume a point-like $\pi N$
four-point coupling that gives rise to the bubble when
$s - M^2 \gg m_\pi^2$.
While a $\delta(x)$ contribution formally appears in the chiral
effective theory framework, this contribution will presumably soften
into a peak near $x=0$ in QCD.  Nevertheless, even if the Adler sum
rule is in the end satisfied in QCD, the presence of $\delta(x)$ terms
in the effective theory indicates that in order to confirm this sum
rule it is essential to include the small-$x$ regime.
Of course, none of these considerations applies to the case of the PS
pion-nucleon coupling, which contains nonzero contributions only at
$y > 0$.

%%%%%%%%%%%%%%%%%%%%%%%%%%%%%%%%%%%%%%%%%%%%%%%%%%%%%%%%%%%%%%%%%%%%%%%%%
\section{Conclusion}
\label{sec:conc}

In this paper we have presented a detailed derivation of the isovector
light-cone momentum distributions of pions in the nucleon for both the
pseudovector and pseudoscalar $\pi N$ interactions.
In the PV theory the direct coupling to the pion gives rise to a pion
rainbow distribution, $f_\pi^{(iv)}$, and a distribution from the pion
bubble, $f_{\pi \rm (bub)}^{(iv)}$, with the latter involving emission
and absorption of a pion at the same point.  Consequently, the bubble 
contribution to the momentum distribution exists only at zero light-cone
momentum, $y=0$.  We found, however, that the pion rainbow diagram also
has a $\delta$-function component, on top of the regular terms at $y>0$.
The $\delta(y)$ contributions arise because the PV $\pi NN$ vertex is
proportional to the momentum of the pion, which leads to a more
singular integrand in the $k^- \to \infty$ limit than in the case
of the momentum-independent PS coupling.

In addition to the pion distribution, we have also computed the
isovector light-cone momentum distribution of the corresponding recoil
nucleon dressed by the pion, including contributions from the rainbow
diagram with nucleon coupling, $f_N^{(iv)}$, and a tadpole coupling
at the $\pi\pi NN$ vertex, $f_{N \rm (tad)}^{(iv)}$, as well as the
Kroll-Ruderman terms $f_{\rm KR}^{(iv)}$ which arise at the same
order in the chiral expansion of the PV Lagrangian.
The latter are in fact essential for preserving the gauge invariance
of the PV theory.  The nucleon rainbow distribution $f_N^{(iv)}$
contains pieces that are associated with the off-shell components
of the nucleon propagator, in addition to the on-shell components,
and a $\delta$-function term analogous to that in the pion light-cone
distribution $f_\pi^{(iv)}$.
In contrast, the nucleon distribution in the PS model only contains
the on-shell contributions.  The KR distribution $f_{\rm KR}^{(iv)}$
has contributions from the off-shell components of the nucleon and
from the singular $\delta$-function terms.
The singular nucleon tadpole distribution is related to the pion
tadpole by
$f_{N \rm (tad)}^{(iv)} = - f_{\pi \rm (bub)}^{(iv)}$ for all $y$.

Combined, the sum of the pion rainbow and KR distributions in the PV
theory is shown to be proportional to the nucleon rainbow distribution,
\begin{eqnarray}
f_\pi^{(iv)}(y)\ +\ f_{\rm KR}^{(iv)}(y) &=& -4 f_N^{(iv)}(y),
\end{eqnarray}
which is a necessary condition for ensuring gauge invariance.
Within the convolution model, we have demonstrated that this in fact
gives the correct number sum rule: the sum of all the pion cloud
contributions does not alter the valence quark number of the nucleon,
although it does of course affect the shape.
This is also true in the PS formulation.

Our results pave the way for future phenomenological applications
of pion cloud models that are manifestly consistent with the chiral
symmetry properties of QCD.  Most calculations in the literature
about the effects of the pion cloud in deep-inelastic scattering have
been made using the PS $\pi N$ coupling \cite{DLY70, Sul72, Spe97},
for which chiral symmetry is not manifest without the addition of
a scalar field.  In fact, the absence of the $\delta$-function
contributions in the PS version of the rainbow diagrams leads to the
coefficient of the leading nonanalytic term in the chiral expansion
of the quark distribution function moments that differs by a factor
4/3 from the PV result \cite{JMT_Z1}.  A detailed study of the
phenomenological consequences of the distributions derived here in
the PV theory will be the subject of an upcoming study \cite{Hen13}.

%%%%%%%%%%%%%%%%%%%%%%%%%%%%%%%%%%%%%%%%%%%%%%%%%%%%%%%%%%%%%%%%%%%%%%%%%
\acknowledgements

This work was supported by the DOE contract No. DE-AC05-06OR23177,
under which Jefferson Science Associates, LLC operates Jefferson Lab,
DOE contract No. DE-FG02-03ER41260, the DOE Science Undergraduate 
Laboratory Internship (SULI) Program, and the Australian Research
Council through an Australian Laureate Fellowship and the ARC Centre
of Excellence in Particle Physics at the Terascale.

%%%%%%%%%%%%%%%%%%%%%%%%%%%%%%%%%%%%%%%%%%%%%%%%%%%%%%%%%%%%%%%%%%%%%%%%%

\end{document}